\documentclass[final,3p,times,twocolumn]{elsarticle}

\usepackage{amssymb}
\usepackage{amsmath}
\usepackage{subcaption}
\usepackage{graphicx}
\usepackage{epstopdf}
\usepackage{physics}
\usepackage{soul}
\usepackage[dvipsnames]{xcolor}
\usepackage{booktabs}
\usepackage{scalerel}
\usepackage{lineno}

\usepackage{hyperref}

\journal{Physics Letters A}

\begin{document}

\begin{frontmatter}

\title{Subharmonic entrainment and limit cycle modulation by high frequency excitation:
A Renormalization group approach} 

\author[label1]{Somnath Roy}
\affiliation[label1]{organization={Institute of Engineering and Management, School of University of Engineering and Management},
            city={Kolkata},
            postcode={700091},
            state={West Bengal},
            country={India}}

\affiliation[label2]{organization={Department of Mathematics},
            addressline={Rammohan College},
            city={Kolkata},
            postcode={700009},
            state={West Bengal},
            country={India}}

\affiliation[label3]{organization={Department of Physics},
            addressline={Jadavpur University},
            city={Kolkata},
            postcode={700032},
            state={West Bengal},
            country={India}}

\author[label2]{Debapriya Das} 

\author[label3]{Dhruba Banerjee}

\begin{abstract}
In this article, we explore the possibility of a sub-harmonic $(1{:}2)$ entrainment, and supercritical Hopf bifurcation in a van der Pol-Duffing oscillator that has been excited by two frequencies, comprising a slow parametric drive and a fast external forcing, through the variation of the amplitude of the external fast signal. We also deduce the condition for the threshold parametric strength required to generate sub-harmonic oscillation. The Blekhman perturbation (direct partition of motion) and the Renormalization group technique have been employed to study how the signal amplitude plays a pivotal role in modulating the limit cycle dynamics as well as the subharmonic generation. Studies of nonlinear responses and bifurcations of such driven nonlinear systems are usually done by treating the strength of the
fast drive as the control parameter. Here we show that, beyond its role in allowing one to study the dynamics with the slow and fast components nicely separated, the amplitude of the high-frequency signal can also be treated as an independent control parameter for controlling both the limit cycle behavior and the onset of subharmonic oscillation in the oscillator. Our analytical estimations are well supported by numerical simulations.

\end{abstract}


\end{frontmatter}



\section{Introduction}

The dynamics of a system subjected to a slow and fast harmonic drive has burgeoned to a significant volume of literature in nonlinear dynamics over the past two decades, and is now commonly addressed as \textit{Vibrational Resonance}. The response of nonlinear systems to two separate driving forces, one with a slower frequency and the other with a frequency much faster than the former, has now become the standard paradigm across physics, biology, engineering, and natural science. A large class of models has been proposed and studied within the ambit of this framework. Among the plethora of works, some pivotal studies have been found in different context including different nonlinear potential systems and physical events \cite{landa2000vibrational,ullner2003vibrational,ghosh2013nonlinear,rajasekar2011novel,jeyakumari2009analysis,rajamani2014ghost}, biological maps and networks \cite{rajasekar2012vibrational,qin2018vibrational,deng2010vibrational}, fault and signal detection \cite{xiao2020weak,xiao2019novel,pan2021study,liu2017detecting}, energy harvesting \cite{coccolo2014energy},non-smooth systems \cite{roy2025vibrational}, plasma \cite{roy2016analysis}, quantum dynamics \cite{roy2025controllingeffectquantumfluctuations} and experimental studies \cite{chizhevsky2003experimental,baltanas2003experimental} as well.An illustrative review in this context can be found in a recent article \cite{yang2024vibrational}. Apart from these, a substantial amount of research has been done concerning different classes of limit cycle oscillators under the architecture of vibrational resonance. The role of these self sustained oscillators arises in diverse fields of science and technology, including biology \cite{pavlidis2012biological}, chemistry \cite {kuramoto2003chemical}, and engineering \cite{teplinsky2008limit}. Effects of external fast forcing on a van der Pol oscillator, a typical representative of limit cycle oscillators, have been explored in several studies \cite{fahsi2009effect,fahsi2009suppression,roy2021vibrational,roy2023controlling} by using perturbation methods like multi-time scale, Lindstedt-Poincaré, and averaging. After the pioneering work of N.Goldenfield \cite{chen1996renormalization}, the renormalization group (RG) perturbation technique has been extensively employed in the analysis of dynamical systems, particularly in the dynamics of limit cycle oscillators \cite{ziane2000certain,sarkar2011center,sarkar2011center,banerjee2010analyzing,banerjee2010renormalization,dutta2018estimating}. Though it has the potential to uncover the fundamental dynamical characteristics of such systems, its use in the context of vibrational resonance is still unexplored.

In the last few decades, studies on the van der Pol oscillator, including Duffing-type nonlinearity, have gained a substantial amount of attention to researchers from diverse fields of application. From engineering to natural science, the implementation can be found in weak signal detection \cite{zhihong2015application}, vibration control \cite{maccari2008vibration,ji2006stability}, machine-learning based modeling \cite{sahoo2024neural}, plasma oscillation \cite{miwadinou2013nonlinear}, biomedical devices \cite{ghorbanian2015phenomenological}, and many more. Beside, a parametric excitation is often considered along with the geometric nonlinearity, which is commonly known as the van der Pol-Mathieu-Duffing (VMD) oscillator. Several studies have been pursued concerning the analytical observations and real-world applications, covering the topics from stability analysis, limit-cycle and quasi-periodic behavior \cite{pandey2007frequency} to energy harvesting using MEMS devices \cite{belhaq2016energy,belhaq2018energy}. 

In a recent work, we have also investigated the VMD oscillator to study vibrational resonance in the presence of an external weak excitation apart from the parametric excitation \cite{roy2021vibrational}. There are two basic motivations behind our studying this oscillator. Firstly, as there is already a parametric frequency associated with the oscillator in addition to its own natural frequency, the study of vibrational resonance for such an oscillator is usually done by bringing in a single forcing term with a very high frequency. Studies pertaining to the response of such systems in the presence of an additional forcing term with a frequency much slower than the fast drive have therefore been rare. Secondly, due to the spatially nonlinear nature of the damping of a self-excited oscillator, it turns out that the action of these twin forcing terms results in a modification of the damping along with the natural frequency of the oscillator. While the former phenomenon, where the natural frequency of the system gets modified in the slow dynamics of the oscillator, is a widely addressed issue in the context of vibrational resonance, the latter phenomenon, where the damping undergoes a modification, is relatively less explored. It is precisely this phenomenon that gives a supercritical Hopf bifurcation in the oscillator that has been studied in our work.

In this article, we have examined the same VMD oscillator with a fast forcing acting on it alone, rather than taking another slow excitation like previous work. This would greatly simplify the problem as the parametric excitation plays the role of slow forcing, preserving the essence of the slow-fast framework. Also, we have extensively found out the role of the parametric threshold on limit cycle oscillation modulated by the fast signal, which has not been investigated earlier. Besides, we have taken the approach of RG perturbation over the multiple scales, cause it offers a more systematic and unified approach to eliminate secular terms using a single time variable, unlike the Multi-Time Scale (MTS) method, which introduces multiple artificial scales. RG naturally captures slow dynamics and invariant manifolds, making it more general and geometrically intuitive.

\section{The model}
The model oscillator that we are studying here is given by the following equation:
\begin{equation}
 \ddot{x} + \gamma(px^2-q) \dot{x} + \omega_0^2 (1 + h \cos \omega_p t)x + \alpha x^3 =  g \cos \Omega t
 \label{eq:original}
\end{equation}

\noindent where $\omega_0$ and $\alpha$ are the linear and nonlinear stiffness of the oscillator. The damping constant related to van der Pol term is given by $\gamma$. The parametric excitation is
done through the term $h \cos \omega_p t$ where the magnitude of the parametric frequency will be determined to account for the parametric strength $h$ across different perturbation orders. A high frequency drive, $g \cos \Omega t$, which is significantly higher than the parametric frequency, $\Omega \gg \omega_0,\omega_p$, is applied to the system on the right side of the equation above. The van der Pol oscillator is known to become entrained at the parametric frequency only if the parametric strength $h$ exceeds a minimum threshold $h_{th}$. The main goal is to identify the corresponding threshold condition that permits sustained subharmonic oscillations because the leading effect of parametric resonance usually appears at half the parametric frequency, i.e., \( \omega_p/2 \). Besides the presence of cubic nonlinearity $\alpha x^3$, which plays a crucial role in modifying the natural frequency of the dynamics, and hence is responsible for the system to oscillate in an effective potential landscape. This effective slow dynamics can be derived by first separating the dynamical variable $x$
into a slow variable $s$ and a fast variable $f$

\begin{equation}
x(t) = s(t,\omega t) + f(t, \Omega t).
\label{eq:seperation}
\end{equation}

\noindent Substituting Eq.\eqref{eq:seperation} into Eq.\eqref{eq:original} we get

\begin{equation}
\begin{split}
\ddot{s}&+\ddot{f}+\gamma\left(p\left(s^2+f^2+2sf\right)-q\right)(\dot{s}+\dot{f})\\
 &+\omega_{0}^{2}(1 + h \cos \omega_p t)(s+f)\\
 &+\alpha(s^3+3s^2 f+3sf^2+f^3)=g\cos(\Omega t).
\end{split}
\label{eq:split_dyn}
\end{equation}

The effective dynamics for the slow variable can be obtained by averaging the above equation over the fast timescale, assuming that the slow variable stays almost constant over the fast time scale, i.e., $\langle x(t)\rangle=\frac{1}{T}\int_{0}^{T}x\,~ \rm d(\Omega t)=s(t,\omega t)$ yields,

\begin{equation}
\begin{split}
\ddot{s}&+\langle\ddot{f}\rangle+\gamma\left(p\left(s^2+\langle f^2\rangle+2s\langle f\rangle\right)-q\right)(\dot{s}+\langle\dot{f}\rangle)\\
&+\omega_{0}^{2}(1 + h \cos \omega_p t)(s+\langle f\rangle)+\alpha(s^3+3s^2\langle f \rangle\\
&+3s\langle f^2\rangle+\langle f^3\rangle)\\
&=\langle g\cos(\Omega t)\rangle.
\label{eq:average}
\end{split}
\end{equation}
The temporal average operator in time period $T=2\pi/\Omega$ is indicated here by $\langle \cdot \rangle$.Now,subtracting Eq.\eqref{eq:average} from Eq.\eqref{eq:split_dyn} and considering the periodicity of the fast variable $\langle\ddot{f}\rangle=\langle\dot{f}\rangle=\langle f\rangle=0$, results in the subsequent equation. 

\begin{equation}
\begin{split}
\ddot{f}&+\gamma p\left((f^2-\langle f^2\rangle)\dot{s}+2s f\dot{s}\right)\\
&+\gamma\left(p(s^2\dot{f}+f^2\dot{f}+2s f\dot{f})-q\dot{f}\right)\\
&+\omega_0^2(1 + h \cos \omega_p t)f+\alpha(f^3-\langle f^3\rangle+3s^2 f\\
&+3s(f^2-\langle f^2\rangle))=g\cos(\Omega t).
\label{eq2.9}
\end{split}
\end{equation}

It is known as the \textit{inertial approximation} \cite{blekhman2000vibrational} since the second derivative of the fast variable dominates the first derivative and the state variable $(i.e.,\ddot{f}>>\dot{f}>f,f^2,f^3...)$. It is now possible to approximate the fast response as by taking the governing equation and reducing it to the simpler form $\ddot{f}\approx g\cos(\Omega t)$.This yields the following solution: $f\approx-\frac{g}{\Omega^2}\cos(\Omega t)=-A_v\cos(\Omega t)$, where $A_v$ is the amplitude of the fast \textit{vibrational signal}. The following statistical properties result from this: $\langle f^2\rangle=\frac{A_v^2}{2}$ represents the power of the signal, and $\langle f\rangle=0$. Eventually, all the average effects of the fast variable get incorporated into the equation
of the slow variable Eq.\eqref{eq:average} thus leading to an effective equation for the slow variable as

\begin{equation}
\ddot{s} + \gamma(ps^2-K) \dot{s} + ({\tilde{\omega}}^2 + \omega_0^2 h \cos \omega_p t)s + \alpha s^3 = 0
\label{eq:effective_dynamics}
\end{equation}

\noindent where the frequency term $\tilde{\omega}$ is given by

\begin{eqnarray}
\tilde{\omega}^2(g) &=& 3\alpha \left\langle f^2 \right\rangle + \omega_0^2
\nonumber \\
&=& \frac{3}{2}\alpha A_v^2+\omega_0^2
\label{eq:effective_freq}
\end{eqnarray}

\noindent and the factor associated with the nonlinear damping $K$ takes the form

\begin{equation}
K(g,\Omega)=q-\frac{g^2}{2\Omega^4}=q-\frac{A_v^2}{2}
\label{eq:effective_damp}
\end{equation}

\noindent Defining an effective frequency as $\omega_{eff}=\sqrt{{\tilde{\omega}}^2+\omega_0^2 h \cos \omega_p t}$ we can identify an effective monostable potential $V_{mono}(s,t)=\frac{1}{2} \omega_{eff}^2 (t) s^2 +\frac{1}{4}\alpha s^4$ while the potential defined by the origianl system Eq.\eqref{eq:original} is \(V(x)=\frac{1}{2}\omega_0^2 x^2+\frac{1}{4}\alpha x^4 \).

\noindent In Eqs.\eqref{eq:effective_freq} and \eqref{eq:effective_damp} we have expressed the effective terms $\tilde{\omega}$
and $K$ as functions of the strength $g$ as well as the high frequency $\Omega$ of the fast driving
force $g\cos\Omega t$. This opens up the scope for studying the consequence of exciting the system
with a slow parametric as well as a fast drive in terms of variation not only of $g$ but also of $\Omega$. The
effect of the fast drive term is twofold. Firstly, from Eqs.(\ref{eq:effective_dynamics}) and (\ref{eq:effective_freq}) we see that the
new frequency $\tilde{\omega}$ emerges as an effective natural frequency thus replacing
$\omega_0$ of Eq.(\ref{eq:original}). Secondly, the Van der Pol damping term gets modified from
$\gamma(px^2 - q)\dot{x}$ in Eq.(\ref{eq:original}) to $\gamma(ps^2 - K)\dot{s}$ of Eq.(\ref{eq:effective_dynamics}). The
question of stability of the limit cycle arises when this effective damping parameter $K$ passes through a
zero in the course of its variation with respect to either $g$ or $\Omega$ or both by $A_v$. Here, we are exploring the role of the fast forcing amplitude ($A_v$) in tuning the stability of the limit cycle.

\section{RG flow at ($1:2$) subharmonic}
\noindent In this work, we investigate the limit cycle behavior entrained by the subharmonic ($1:2$) parametric frequency. To make further progress, we need to derive the amplitude and phase flow equations from
Eq.(\ref{eq:effective_damp}) with the parametric frequency replaced by a value that is twice the value of the slow
driving frequency, i.e., $\omega_p = 2\tilde{\omega}$. The flow equations can be arrived at
through some standard perturbation techniques, for example, multiple-time-scale analysis, averaging, harmonic balance, etc. Here we adapt well known RG method to obtain the flow equations. Defining the dimensionless time $\tau = \omega_p t$, we
do the perturbation around the redressed natural frequency $\tilde{\omega}$ by making the
frequency $\omega_p$ of the slow forcing drive perturbatively close to $2\tilde{\omega}$ through the
introduction of a detuning $\tilde{\sigma}$ as $\omega_p = 2\tilde{\omega} + \delta \tilde{\sigma}$,
where $\delta$ is the perturbation parameter denoting smallness. Introducing the rescaled
parameters $\eta = \gamma / \omega_p$, $\beta = \alpha / {\omega_p}^2$, and $\epsilon = h{\omega_0^2} / {\omega_p}^2$, we can rearrange Eq.(\ref{eq:effective_dynamics}) as

\begin{equation}
s'+\frac{1}{4}s = -\eta (ps^2 - K)s'- \epsilon (\cos \tau) s - \beta s^3 +\delta \sigma
s
\label{eq:effective_redressed}
\end{equation}

\noindent where the prime ($'$) denotes the derivative with respect to $\tau$ and the rescaled detuning parameter is defined to first order in $\delta$ as
${\tilde{\omega}}^2 = \omega_p^2 (\frac{1}{4} - \delta \sigma)$,where $\sigma=\frac{\tilde{\omega}\tilde{\sigma}}{\omega_p^2}$. For using RG analysis, we can
write the perturbation expansion to the $1^{st}$ order of the perturbation parameters $\eta,\delta,\epsilon,\beta$ as:
\begin{equation}
s(\tau) = s_0(\tau) + \eta s_{1\eta} +\delta s_{1\delta}+\epsilon s_{1\epsilon}+\beta s_{1\beta}+\text{higher orders}\cdots
\label{eq:power_series}
\end{equation}
and by substituting Eq.\eqref{eq:power_series} into Eq.\eqref{eq:effective_redressed} we obtain the following hierarchy of equations by equating the successive powers of perturbation parameters as :

\begin{eqnarray}
    s_0''+\frac{1}{4}s_0=0~~~:(\text{zeroth order})\label{eq:0th_order}\\
    s_{1\eta}''+\frac{1}{4}s_{1\eta}=-(ps_0^2-K)s_0'~~~:\mathcal{O}(\eta^1)\label{eq:1storder_eta}\\
    s_{1\delta}''+\frac{1}{4}s_{1\delta}=\sigma s_0~~~:\mathcal{O}(\delta^1)\label{eq:1storder_delta}\\
    s_{1\epsilon}''+\frac{1}{4}s_{1\epsilon}=-\cos(\tau)s_0~~~\mathcal{O}(\epsilon^1)\label{eq:1storder_epsilon}\\
    s_{1\beta}''+\frac{1}{4}s_{1\beta}=-s_0^3~~~\mathcal{O}(\beta^1)\label{eq:1storder_beta}
\end{eqnarray}

The zeroth order solution yields:

\begin{equation}
s_0=a\cos(\frac{\tau}{2})
\label{eq:0th_sol}
\end{equation}

where we take the initial condition $s_0(\tau=0)=a$ and $\dot{s}_0(\tau=0)=0$.Also, the subsequent initial conditions for higher order terms are chosen as $s_1(0)=\dot{s}_1(0)=0$. using Eq.\eqref{eq:0th_sol} and the initial conditions, Eq.\eqref{eq:1storder_eta} yields the solution

\begin{equation}
\begin{split}
s_{1\eta}&=\frac{pa^3}{16}\left(3\sin\frac{\tau}{2}-\sin\frac{3\tau}{2}\right)+\frac{\tau}{2}\left(K-\frac{pa^2}{4}\right)a\cos\frac{\tau}{2}\\
&-\frac{1}{2}\left(2K-\frac{pa^2}{2}\right)a\sin\frac{\tau}{2}
\end{split}
\label{eq:1st_eta_sol}
\end{equation}
Similarly, Eq.\eqref{eq:1storder_delta} gives

\begin{equation}
    \begin{split}
        s_{1\delta}=\sigma a \tau\sin\frac{\tau}{2}
    \end{split}
    \label{eq:1st_delta_sol}
\end{equation}
The solution of Eq.\eqref{eq:1storder_epsilon} reads

\begin{equation}
    \begin{split}
        s_{1\epsilon}=-\frac{a}{4}\cos\frac{\tau}{2}+\frac{a}{4}\cos\frac{3\tau}{2}-\frac{a}{2}\tau\sin\frac{\tau}{2},
    \end{split}
    \label{eq:1st_epsilon_sol}
\end{equation}

and by Eq.\eqref{eq:1storder_beta} we get

\begin{equation}
    \begin{split}
        s_{1\beta}=-\frac{\beta a^3}{8}\cos\frac{\tau}{2}-\frac{3\beta a^3}{4}\tau\sin\frac{\tau}{2}+\frac{\beta a^3}{8}\cos\frac{3\tau}{2}
    \end{split}
    \label{eq:1st_beta_sol}
\end{equation}
Combining Eqns. \eqref{eq:0th_sol}-\eqref{eq:1st_beta_sol} and put it back into Eq.\eqref{eq:power_series}, we obtain the expression for $s$ up to first order correction as:

\begin{equation}
    \begin{split}
        s(\tau)&=a\cos\frac{\tau}{2}+\eta\Bigg[\frac{pa^3}{16}\left(3\sin\frac{\tau}{2}-\sin\frac{3\tau}{2}\right)\\
&+\frac{\tau}{2}\left(K-\frac{pa^2}{4}\right)a\cos\frac{\tau}{2}-\frac{1}{2}\left(2K-\frac{pa^2}{2}\right)a\sin\frac{\tau}{2}\Bigg]\\
&+\sigma a \tau\sin\frac{\tau}{2}+\epsilon\Bigg[\frac{a}{4}\left(\cos\frac{3\tau}{2}-\cos\frac{\tau}{2}\right)-\frac{a\tau}{2}\sin\frac{\tau}{2}\Bigg]\\
&+\beta\Bigg[-\frac{a^3}{8}\left(\cos\frac{\tau}{2}-\cos\frac{3\tau}{2}\right)-\frac{3a^3 \tau}{4}\sin\frac{\tau}{2}\Bigg]\\
&+\text{higher orders}\cdots
    \end{split}
    \label{eq:s_complete_sol}
\end{equation}

At this point, by taking advantage of the freedom in selecting the initial point $\tau=0$, the renormalization group method can produce an infinite set of time scales. This is accomplished by introducing an arbitrary time scale $\mu$ \cite{banerjee2010analyzing}, and expressing $\tau$ in terms of $\mu$ in the divergent part of Eq.\eqref{eq:s_complete_sol} by splitting it as $[\tau-\mu]+[\mu-0]$. To continue, we keep in mind that the first set of brackets represents the ``present'' whereas the later includes contributions from the ``past''. In order to push all divergent behavior into the past, we employ the arbitrary time scale \(\mu\) somewhere in the interval \(0 \rightarrow t\) in such a way that the present time scale will be divergent-free. At firs,t let us renormalize the initial amplitude and phase with the renormalization factors as

\begin{equation}
\begin{split}
    a=a_0=a(\mu)Z_1(\mu)&=a(\mu)\Big(1+\eta Z_{1\eta}+\delta Z_{1\delta}\\
    &+\epsilon Z_{1\epsilon}+\beta Z_{1\beta}+\cdots\Big)
 \end{split}
 \label{eq:renor_amp}
\end{equation}

\begin{equation}
\begin{split}
0=\theta_0=\theta(\mu)+Z_2(\mu)&=\theta(\mu)+\eta Z_{2\eta}+\delta Z_{2\delta}\\
    &+\epsilon Z_{2\epsilon}+\beta Z_{2\beta}+\cdots
\end{split}
\label{eq:renor_phase}
\end{equation}    
Here, $Z_{1i}$ and $Z_{2i}$, where $i=\eta,\delta,\epsilon,\beta$ are the arbitrary constants have to be determined order by order to remove the divergence terms.The renormalized quantities \(a\) and \(\theta\), which correlate to the parameters \(a_0\) and \(\theta_0\), respectively, are then obtained by absorbing the \(\mu\)-dependent terms. Specifically, \(a\) and \(\theta\) now explicitly depend on \(\mu\). In light of this, we can rewrite the Eq.\eqref{eq:s_complete_sol} by using Eqns.\eqref{eq:renor_amp} and \eqref{eq:renor_phase} in terms of the coeeficients of $\sin\frac{\tau}{2}$ and $\cos\frac{\tau}{2}$:

\begin{multline}
        s(\tau,\mu)=a\cos\phi+\frac{p\eta a^3}{16}\left(3\sin\phi-\sin3\phi\right)-\\\frac{\eta}{2}\left(2K-\frac{pa^2}{2}\right)a\sin\phi
        +\eta a\cos\phi\Bigg[ Z_{1\eta}\\ \left.+\frac{1}{2}\left(K-\frac{pa^2}{4}\right)(\tau-\mu)+\frac{\mu}{2}\left(K-\frac{pa^2}{4}\right)  \right]
        +\\ \delta a\sin\phi\left[Z_{2\delta}+\sigma(\tau-\mu)+\sigma\mu\right]+\frac{\epsilon a}{4}(\cos3\phi-\cos\phi)\\
        -\epsilon a\sin\phi\left[Z_{2\epsilon}+\frac{1}{2}(\tau-\mu)+\frac{\mu}{2}\right]-\frac{\beta a^3}{8}(\cos\phi-\cos3\phi)\\
        +\beta a\sin\phi\left[Z_{2\beta}-\frac{3a^2}{4}(\tau-\mu)-\frac{3a^2\mu}{4}\right],
        \label{eq:renormalized_eq}
\end{multline}
where we use the expansion up to first order

\begin{equation}
\begin{split}
    &a\cos\phi=a(\mu)\left(1+\eta Z_{1\eta}+\delta Z_{1\delta}+\epsilon Z_{1\epsilon}+\beta Z_{1\beta}\right)\cos\phi\\
    &-\left(\eta Z_{2\eta}+\delta Z_{2\delta}+\epsilon Z_{2\epsilon}+\beta Z_{2\beta}\right)\sin\phi
    \end{split}
    \label{eq:cosphi_def}
\end{equation} with, $\phi(\mu) = \frac{\tau}{2} + \theta(\mu)$.
Now, we can define the \( Z \)-functions in a way that cancels out the divergences that show up in the \( [\mu-0] \) terms. In accordance with the renormalization group prescription, these disparate contributions can subsequently be assimilated into the ``past." Then the natural choice comes at the different orders as: $Z_{1\eta}=-\frac{\mu}{2}\left(K-\frac{pa^2}{4}\right),Z_{2\eta}=0,Z_{1\delta}=0,Z_{2\delta}=-\sigma\mu,Z_{1\epsilon}=0,Z_{2\epsilon}-\frac{\mu}{2},Z_{1\beta}=0,\text{and}~Z_{2\beta}=\frac{3a^2\mu}{4}$. Now the $(\tau-\mu)$ terms can absorbed in the renormalized $a(\mu)$ and $\phi(\mu)$ so that Eq.\eqref{eq:renormalized_eq} yields:

\begin{equation}
    \begin{split}
        s(\tau,\mu)&=a\cos\phi+\frac{p\eta a^3}{16}\left(3\sin\phi-\sin3\phi\right)\\
        &-\frac{\eta}{2}\left(2K-\frac{pa^2}{2}\right)a\sin\phi\\
        &+\eta a\cos\phi\left[\frac{1}{2}\left(K-\frac{pa^2}{4}\right)(\tau-\mu)\right]\\
        &+\delta a\sin\phi\left[\sigma(\tau-\mu)\right]+\frac{\epsilon a}{4}(\cos3\phi-\cos\phi)\\
        &-\epsilon a\sin\phi\left[\frac{1}{2}(\tau-\mu)\right]-\frac{\beta a^3}{8}(\cos\phi-\cos3\phi)\\
        &+\beta a\sin\phi\left[\frac{3a^2}{4}(\tau-\mu)\right],
        \end{split}
        \label{eq:renormalized_eq1}
\end{equation}

As the arbitrary scale \( \mu \) does not appear in the original dynamics, one expects that Eq.~\eqref{eq:renormalized_eq1} should be independent of \( \mu \). independent of $\mu$, such that \( \frac{\partial s}{\partial \mu} = 0 \), which leads to the RG flow of amplitude and phase as:

\begin{eqnarray}
   \frac{\partial a}{\partial \mu}=\frac{\eta a}{2}\left(K-\frac{pa^2}{4}\right)\label{eq:rg_flow_amp}\\
   \frac{\partial \phi}{\partial \mu}=\frac{3}{4}\beta a^2-\delta \sigma+\frac{\epsilon}{2}\label{eq:rg_flow_phase}
\end{eqnarray}

Finally, as $\mu$ is arbitrary we can set $\mu=\tau=\omega_p t$, that guides to the solution of Eq.\eqref{eq:renormalized_eq1} up to first order 

\begin{equation}
    \begin{split}
         s(\tau,\mu)&=a\cos(\Theta t+\theta_0)+\frac{p\eta a^3}{16}\Big(3\sin(\Theta t+\theta_0)\\
         &-\sin3(\Theta t+\theta_0)\Big)
        -\frac{\eta}{2}\left(2K-\frac{pa^2}{2}\right)a\sin(\Theta t+\theta_0)\\
        &+\frac{\epsilon a}{4}(\cos3(\Theta t+\theta_0)-\cos(\Theta t+\theta_0))\\
        &-\frac{\beta a^3}{8}(\cos(\Theta t+\theta_0)-\cos3(\Theta t+\theta_0)),\\
    \end{split}
    \label{eq:final_slow_sol}
\end{equation}

where the corrected frequency is given by

\begin{equation}
    \Theta=\left(\frac{1}{2}+\frac{3}{4}\beta a^2-\delta\sigma-\frac{\epsilon }{2}\right)\omega_p
    \label{eq:corrected_freq}
\end{equation}

\section{Parametric threshold and Hopf-induced limit cycle}

Now, the fixed points of Eqns. \eqref{eq:rg_flow_amp} and \eqref{eq:rg_flow_phase}, obtained by putting the derivatives equal to zero lead
us to those specific locations in the parameter space where we can get the nonlinear responses of the system depicted by peaks in the amplitude curve, which can be expressed by the equation:

\begin{equation}
    \left(\frac{\eta}{2}\left(K-\frac{pa^2}{4}\right)\right)^2+\left(\frac{3}{4}\beta a^2-\delta \sigma+\frac{\epsilon}{2}\right)^2=0.
    \label{eq:response_curve}
\end{equation}
The above equation becomes quadratic in \( \rho \) if \( a = \rho^2 \) is assumed, leads to
\begin{equation}
    \begin{split}
        \left(\frac{\eta^2 p^2}{64}+\frac{9\beta^2}{16}\right)\rho^2&-\left(\frac{pK\eta^2}{8}+\frac{3\beta\delta\sigma}{2}\right)\rho\\
        &\left(\delta^2\sigma^2+\frac{\eta^2K^2}{4}-\frac{\epsilon^2}{4}\right)=0
    \end{split}
    \label{eq:rho_eqn}
\end{equation}

This enables us to solve for the amplitude explicitly in terms of the parameters of the system as:

\begin{equation}
    \rho=\frac{(2p\eta^2 K+24\beta\delta\sigma)\pm\sqrt{\Delta}}{\left(18\beta^2+\frac{\eta^2 p^2}{32}\right)}
    \label{eq:amlitude_sol},
\end{equation}

where the discriminant is
\begin{equation}
    \Delta=\epsilon^2\left(\frac{9}{4}\beta^2+\frac{\eta^2 p^2}{16}\right)-\frac{\eta^2}{4}\left(p\delta\sigma-3\beta K\right)^2\label{eq:discriminant}
\end{equation}
By which we can argue that the positivity of the discriminant $\Delta>0$ confirms the real root of the Eq.\eqref{eq:rho_eqn}. Now Eq.\eqref{eq:rho_eqn} has two positive roots if 
\begin{equation}
    \begin{split}
        &\left(\frac{pK\eta^2}{8}+\frac{3\beta\delta\sigma}{2}\right)>0\\
    &\left(\delta^2\sigma^2+\frac{\eta^2K^2}{4}-\frac{\epsilon^2}{4}\right)>0
    \end{split}
    \label{eq:two_positive_root}
\end{equation}
and has only one positive root if 
\begin{equation}
   \left(\delta^2\sigma^2+\frac{\eta^2K^2}{4}-\frac{\epsilon^2}{4}\right)<0.
   \label{eq:one_positive_root}
\end{equation}
Eq.\eqref{eq:one_positive_root} guarantees the existence of a single limit cycle oscillation at sub-harmonic yields the condition for the threshold of parametric strength as expressed in terms of the original parameters

\begin{figure}[h]
    \includegraphics[scale=0.55]{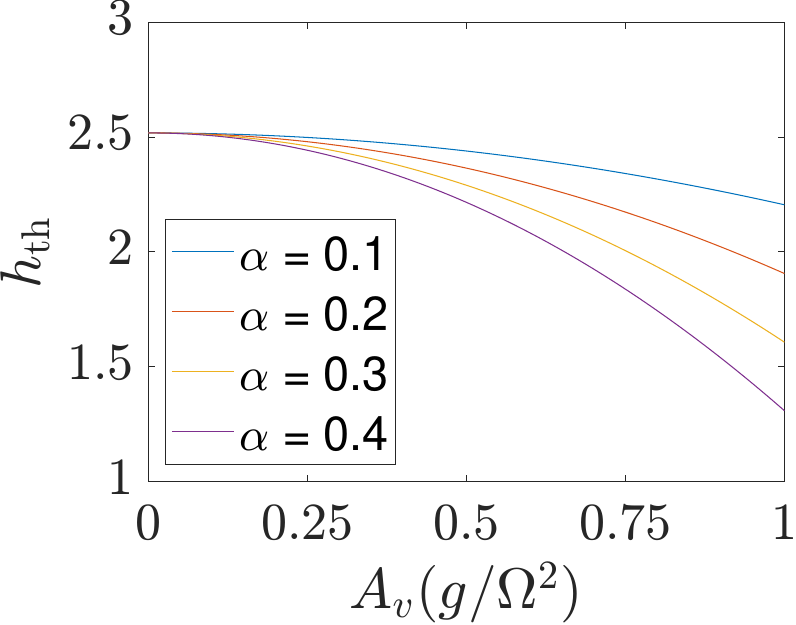}
    \caption{Variation of threshold parametric forcing strength $h_{th}$ with the fast signal amplitude $A_v$ for different values of nonlinear coefficients $\alpha$. With increasing signal strength, the threshold gets diminished, showing greater susceptibility to parametric excitation.}
    \label{fig:h_th_Av}
\end{figure}

\begin{figure*}[h]
    \subcaptionbox{$h=0.1,A_v=1.35$}{\includegraphics[scale=0.4]{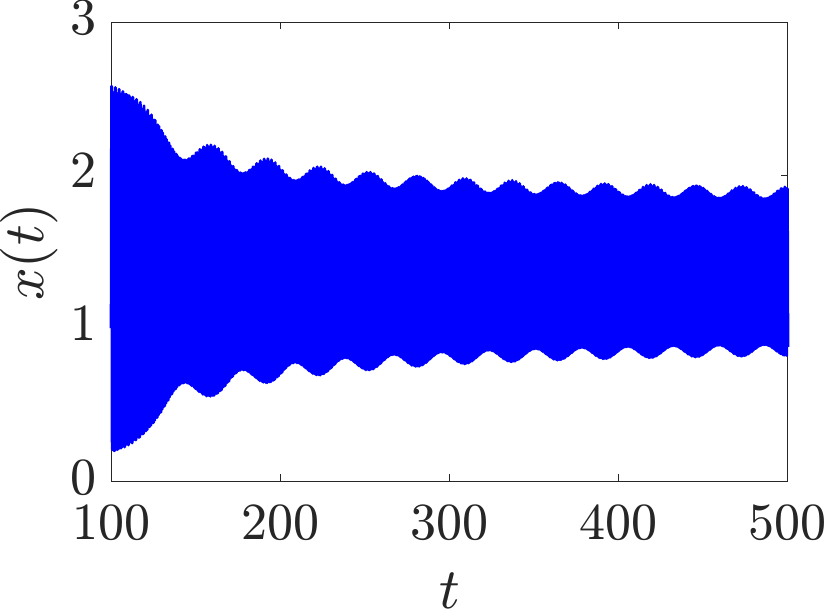}}
    \subcaptionbox{$h=0.1,A_v=1.35$}{\includegraphics[scale=0.4]{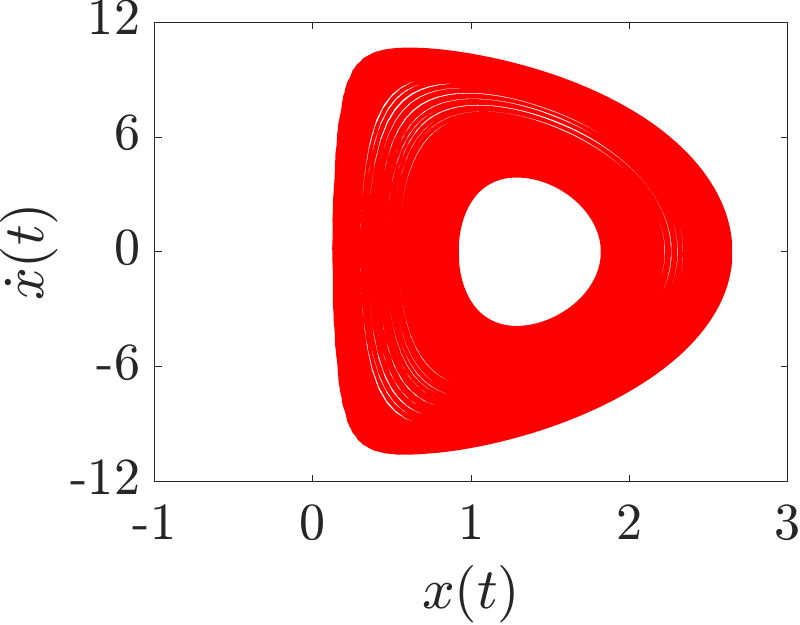}}
    \subcaptionbox{$h=0.1,A_v=1.35$}{\includegraphics[scale=0.4]{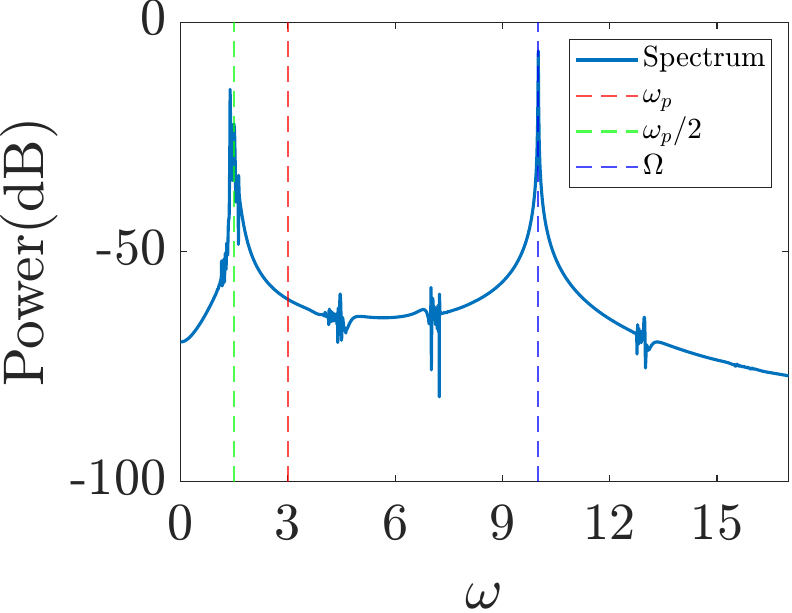}}
        
    \subcaptionbox{$h=0.1,A_v=1.35$}{\includegraphics[scale=0.4]{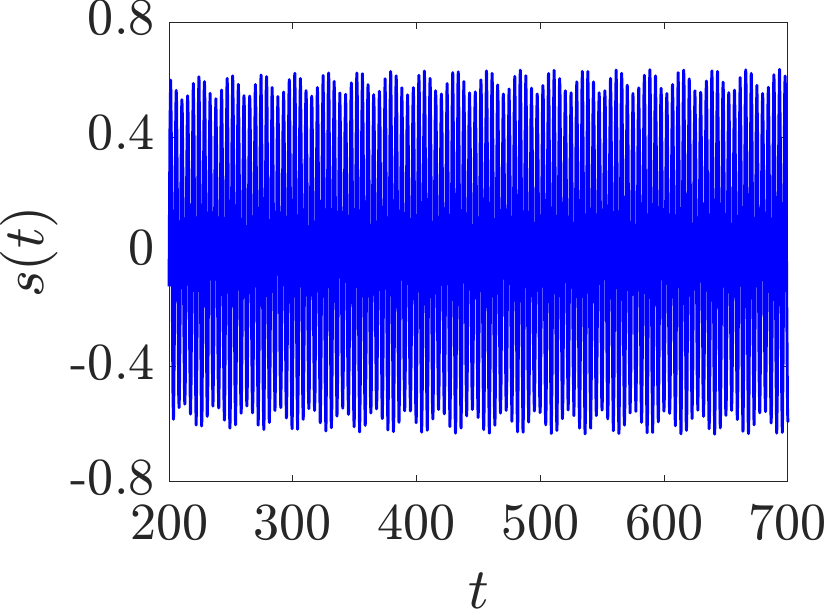}}
    \subcaptionbox{$h=0.1,A_v=1.35$}{\includegraphics[scale=0.4]{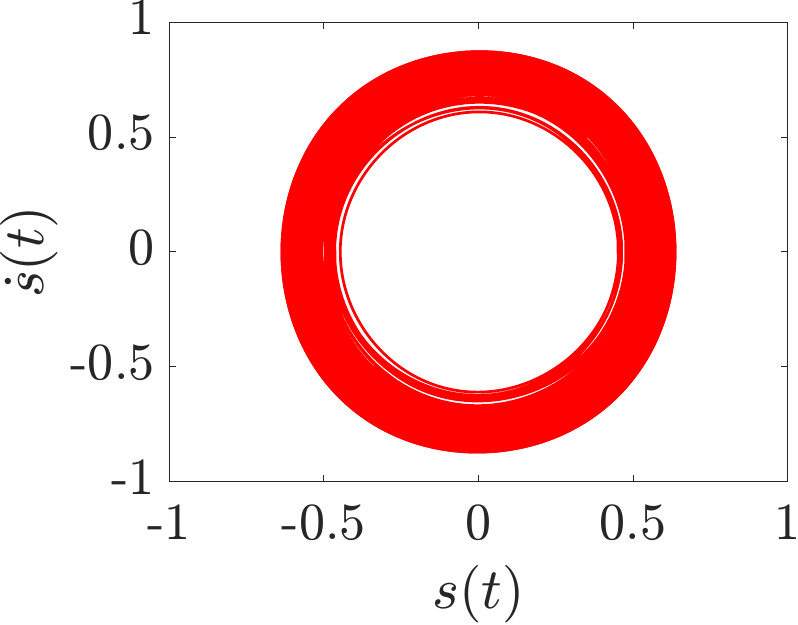}}
    \subcaptionbox{$h=0.1,A_v=1.35$}{\includegraphics[scale=0.4]{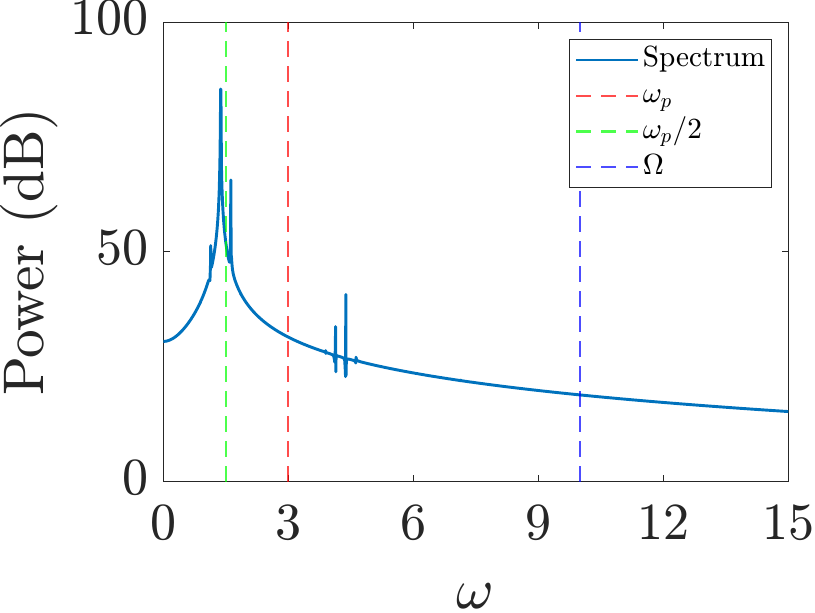}}
\caption{Limit cycle oscillation of the system when \(h<h_{th}\) and \(A_v<(A_v)_{hopf}\). The top row describes the original dynamics Eq.\eqref{eq:original} and the bottom row represents the effective dynamics Eq.\eqref{eq:effective_dynamics}. Each column corresponds to: (a,d) for time series, (b,e) for phase portrait, and (c,f) for the power spectral density (FFT) of the respective dynamics. It can be observed that a clear formation of a limit cycle around $\omega_p/2$ is accompanied by mixed frequency components. The other parameters are taken as: $\alpha=0.3,\gamma=0.1,p=1,q=1,\omega_0=1,\omega_p=3.0$ and $g=135,\Omega=10$.}
\label{fig1}
\end{figure*}

\begin{equation}
    h_{th}=\frac{1}{\omega_0^2}\sqrt{4\left(\tilde{\omega}^2-\frac{\omega_p^2}{4}\right)^2+\gamma^2\omega_p^2K^2},
    \label{eq:threshold_para_strength}
\end{equation}

where $\tilde{\omega}$ and $K$ are given by Eqns. \eqref{eq:effective_freq} and \eqref{eq:effective_damp}. It is immediately apparent from Eq.\eqref{eq:threshold_para_strength} that the threshold is controlled only by the nonlinear damping term \( x^2 \) when the cubic nonlinearity is absent. However, the combined effect of the modified natural frequency \( \tilde{\omega} \) and the effective damping parameter \( K \) can lower the threshold even further in the presence of the cubic stiffness parameter \( \alpha \). As the signal strength \( A_v \) determines both \( \tilde{\omega} \) and \( K \), we have depicted the variation of $h_{th}$ with it in Fig.\ref{fig:h_th_Av}. Consequently, raising \( A_v \) lowers the threshold for parametric instability. But we need to take a moment to think about a subtle point. It is clear from looking at the structure of the amplitude flow equations Eq.\eqref{eq:rg_flow_amp}, that \( A_v \) also affects \( K \), which in turn affects the stability of the limit cycle. This raises the possibility that the nonlinear coupling caused by the fast forcing could result in the emergence of a large-amplitude limit cycle oscillation under the right circumstances.\\

\begin{figure*}[t]
    \subcaptionbox{$h=0.1,A_v=1.5$}{\includegraphics[scale=0.4]{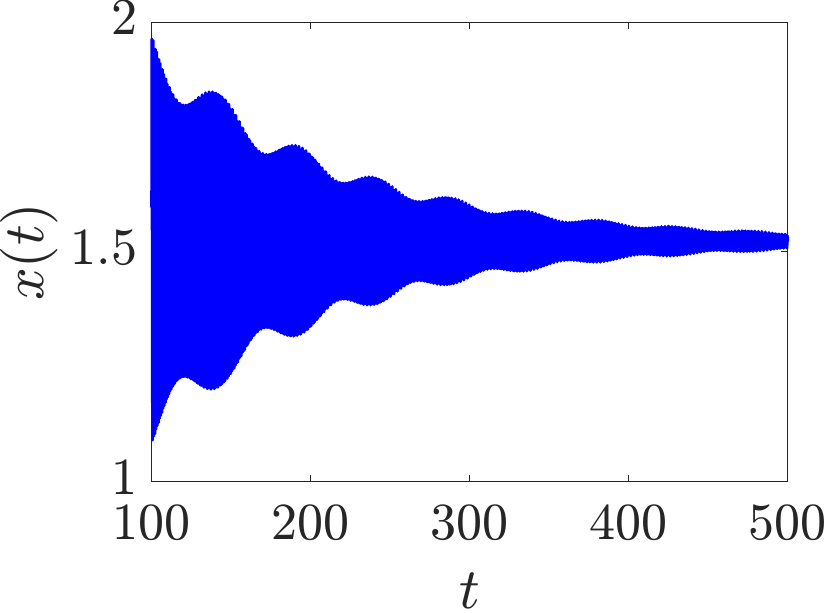}}
    \subcaptionbox{$h=0.1,A_v=1.5$}{\includegraphics[scale=0.4]{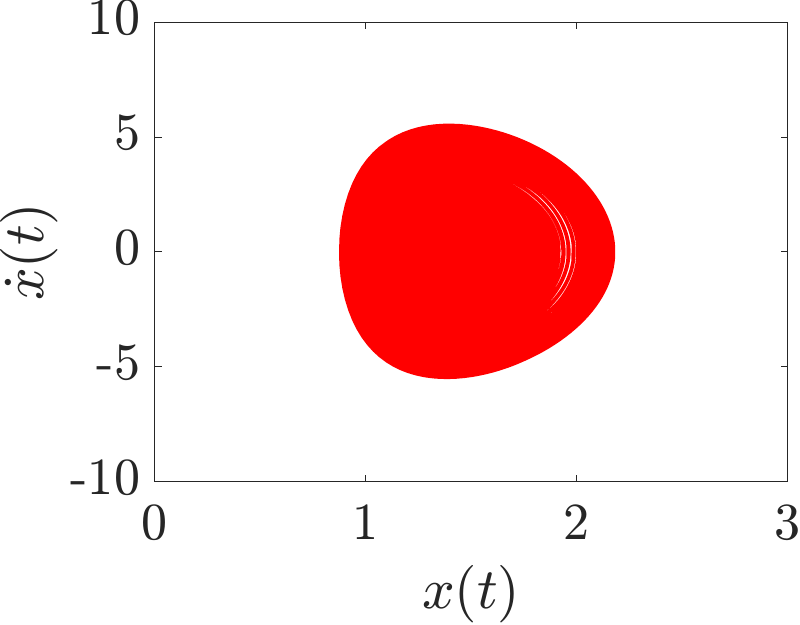}}
    \subcaptionbox{$h=0.1,A_v=1.5$}{\includegraphics[scale=0.4]{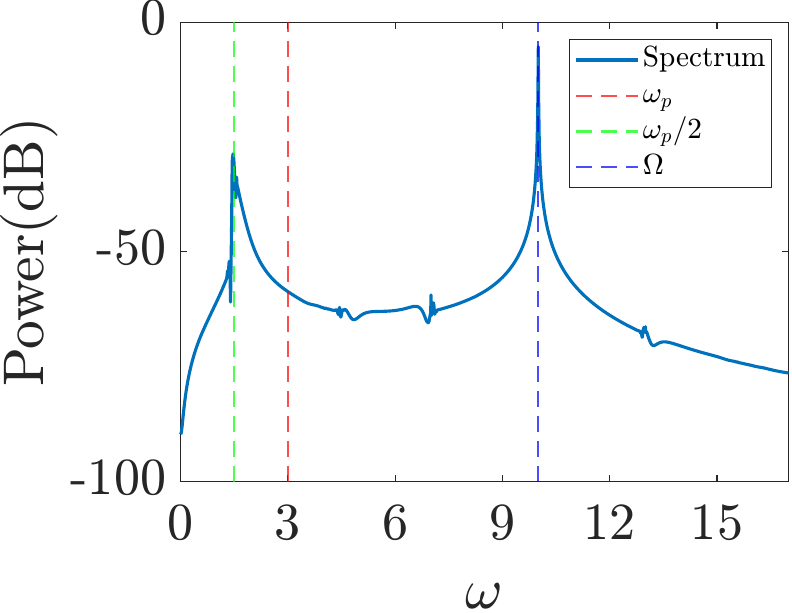}}
        
    \subcaptionbox{$h=0.1,A_v=1.5$}{\includegraphics[scale=0.4]{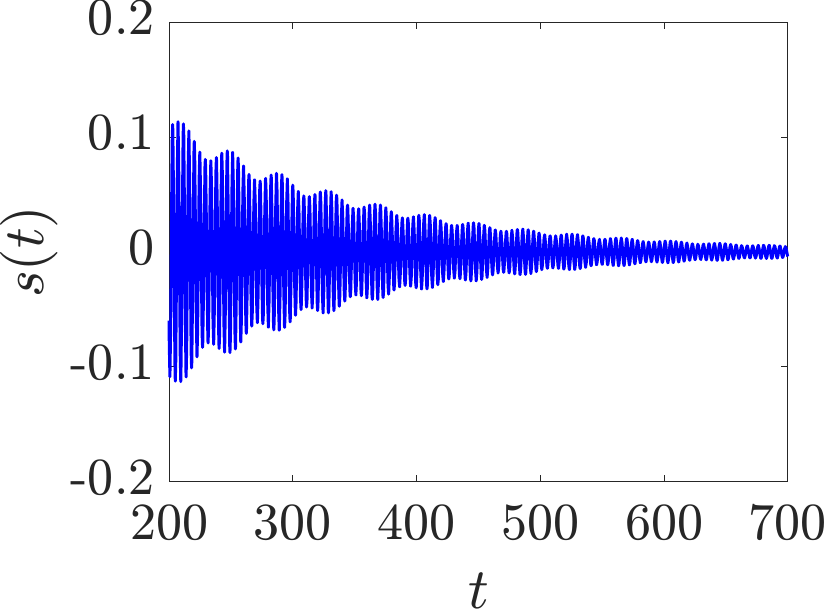}}
    \subcaptionbox{$h=0.1,A_v=1.5$}{\includegraphics[scale=0.4]{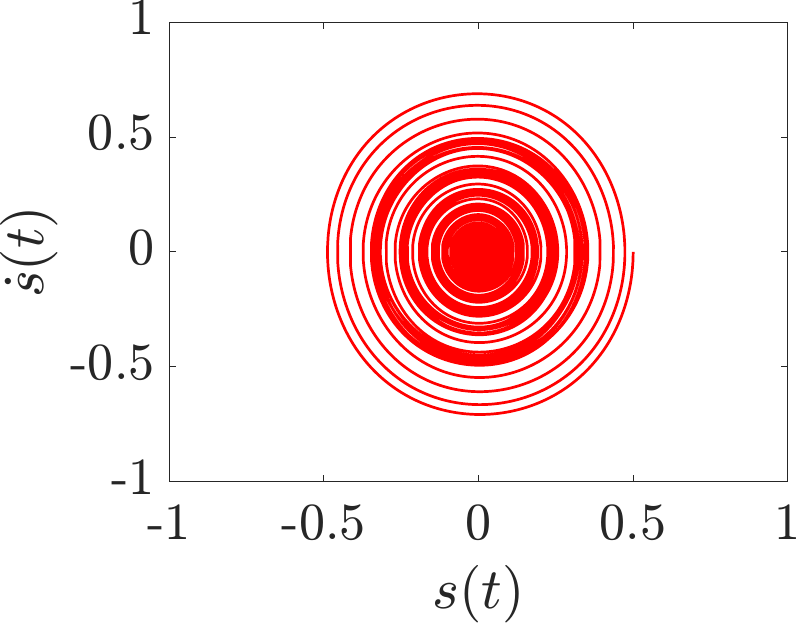}}
    \subcaptionbox{$h=0.1,A_v=1.5$}{\includegraphics[scale=0.4]{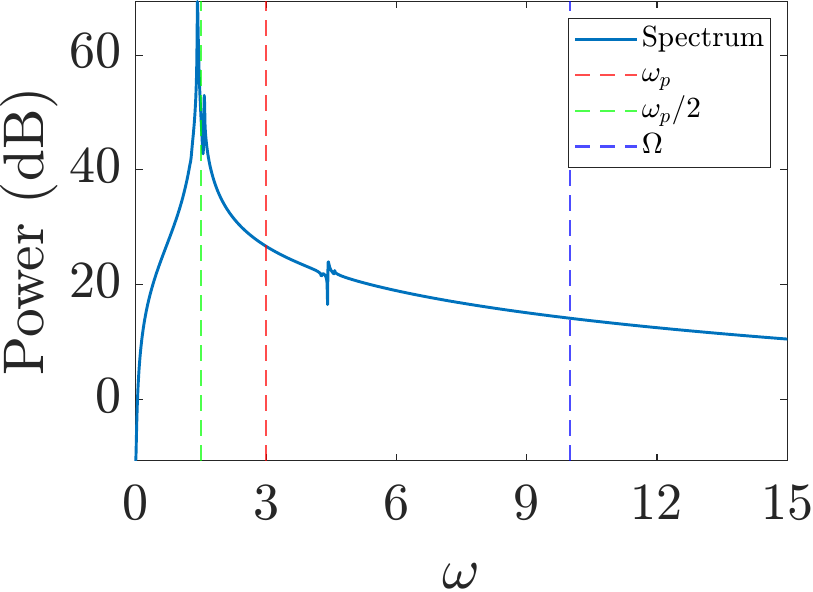}}
\caption{Collapse of the limit cycle to a fixed point by supercritical Hopf as \(A_v>(A_v)_{hopf}\) with $h<h_{th}$. The top row describes the original dynamics Eq.\eqref{eq:original} and the bottom row represents the effective dynamics Eq.\eqref{eq:effective_dynamics}. Each column corresponds to: (a,d) for time series, (b,e) for phase portrait, and (c,f) for the power spectral density (FFT) of the respective dynamics. Once the amplitude of the fast signal exceeds a critical threshold, it is apparent that the system ceases to oscillate gradually. The other parameters are taken as: $\alpha=0.3,\gamma=0.1,p=1,q=1,\omega_0=1,\omega_p=3.0$ and $g=150,\Omega=10$.}
\label{fig2}
\end{figure*}

\begin{figure*}[t]
    \subcaptionbox{$h=0.9,A_v=1.5$}{\includegraphics[scale=0.4]{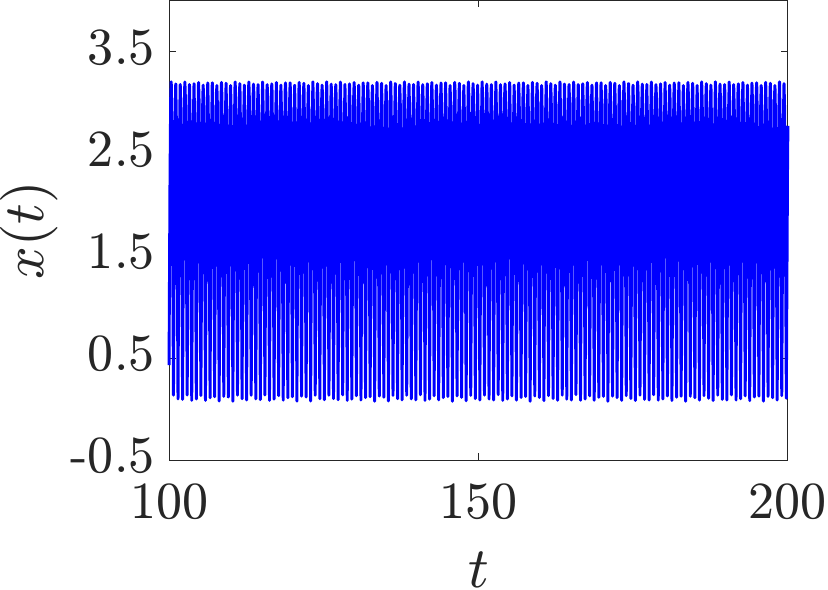}}
    \subcaptionbox{$h=0.9,A_v=1.5$}{\includegraphics[scale=0.4]{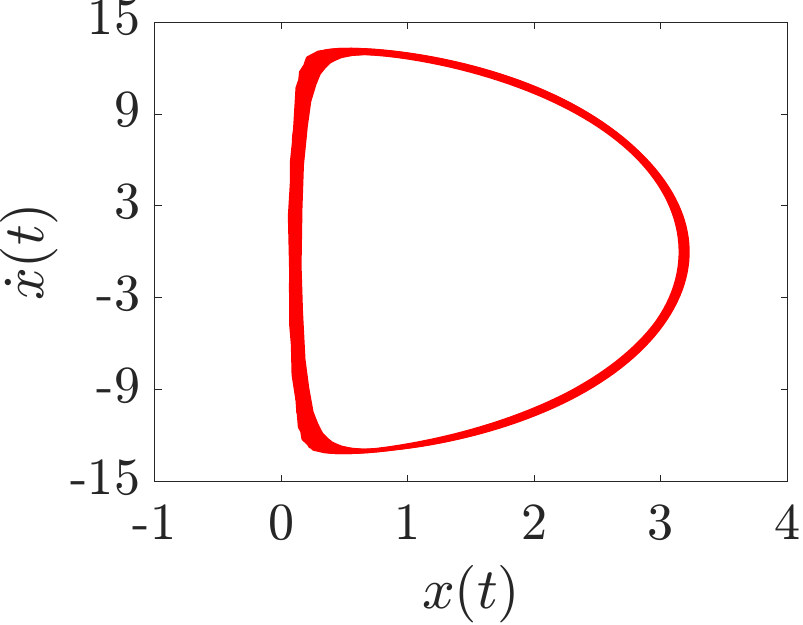}}
    \subcaptionbox{$h=0.9,A_v=1.5$}{\includegraphics[scale=0.4]{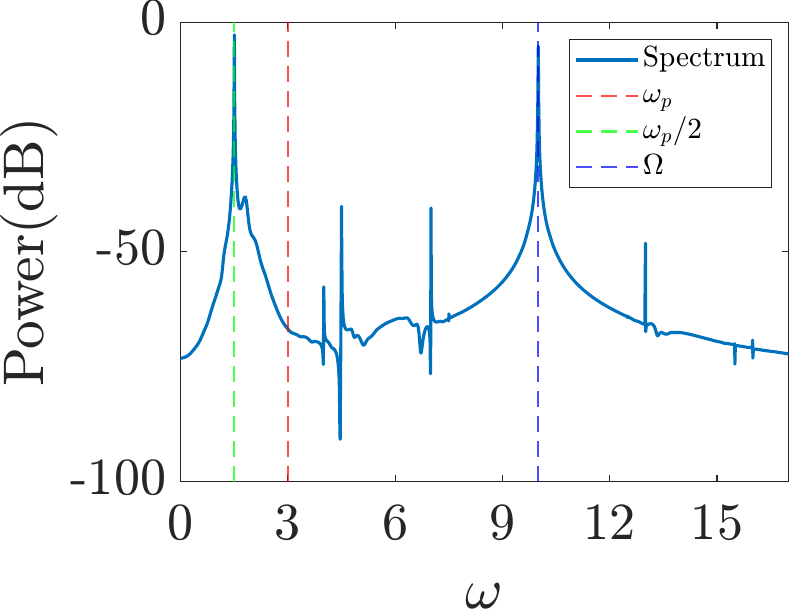}}
        
    \subcaptionbox{$h=0.9,A_v=1.5$}{\includegraphics[scale=0.4]{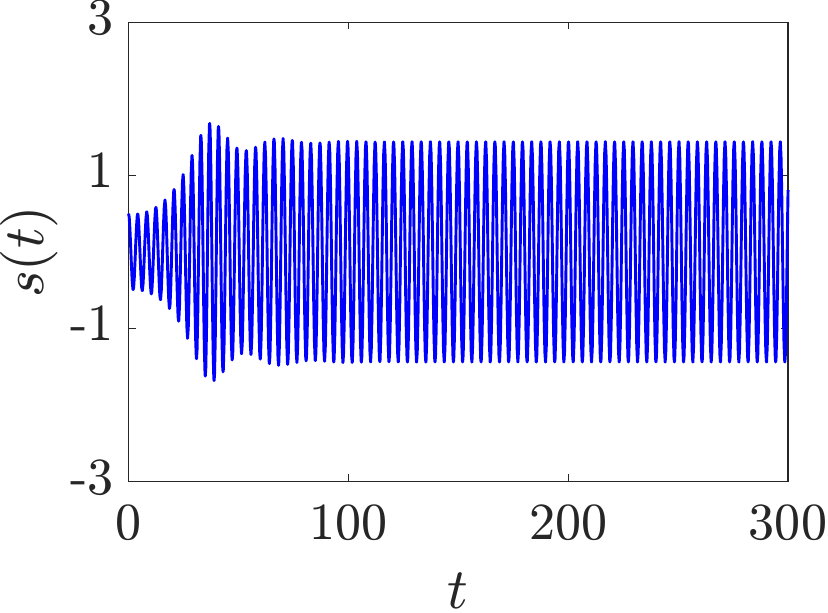}}
    \subcaptionbox{$h=0.9,A_v=1.5$}{\includegraphics[scale=0.4]{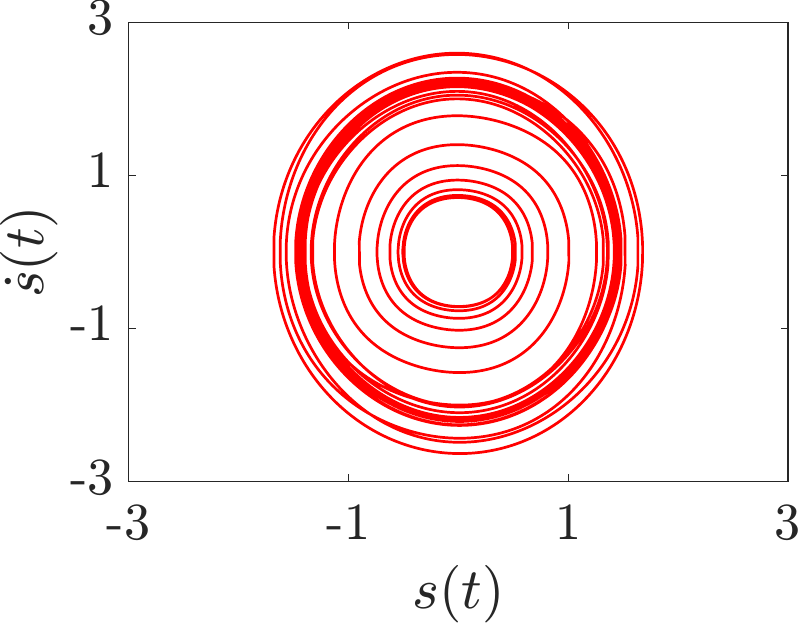}}
    \subcaptionbox{$h=0.9,A_v=1.5$}{\includegraphics[scale=0.4]{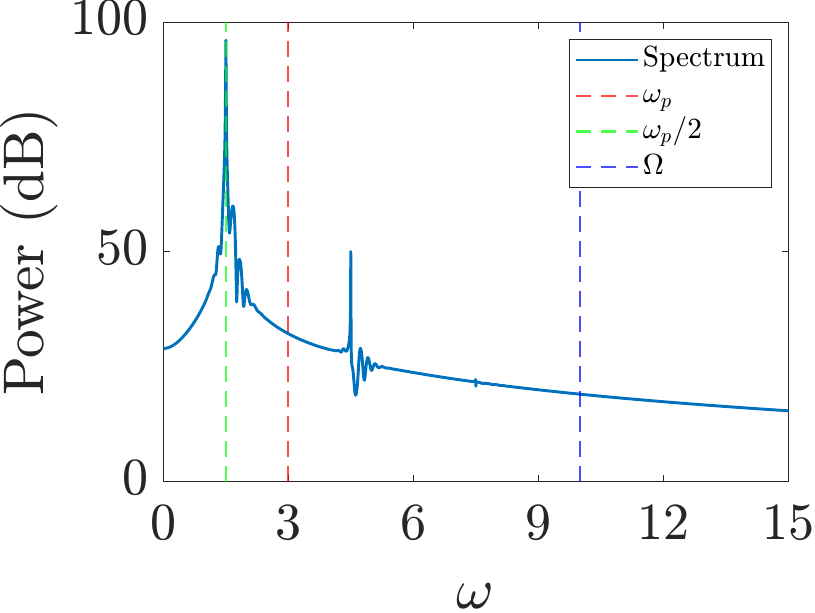}}
\caption{Sustained subharmonic oscillation when $h>h_{th}$. Here $h=0.9$ and $h_{th}\approx 0.5$ from Eq.\eqref{eq:threshold_para_strength} for the given set of parameters. The top row describes the original dynamics Eq.\eqref{eq:original} and the bottom row represents the effective dynamics Eq.\eqref{eq:effective_dynamics}. Each column corresponds to: (a,d) for time series, (b,e) for phase portrait, and (c,f) for the power spectral density (FFT) of the respective dynamics. Once the amplitude of the fast signal exceeds a critical threshold, it is apparent that the system ceases to oscillate gradually. The other parameters are taken as: $\alpha=0.3,\gamma=0.1,p=1,q=1,\omega_0=1.0,\omega_p=3.0$ and $g=150,\Omega=10$.}
\label{fig:h_gt_h_th}
\end{figure*}

\begin{figure}[h]
     \subcaptionbox{Time series}{\includegraphics[scale=0.45]{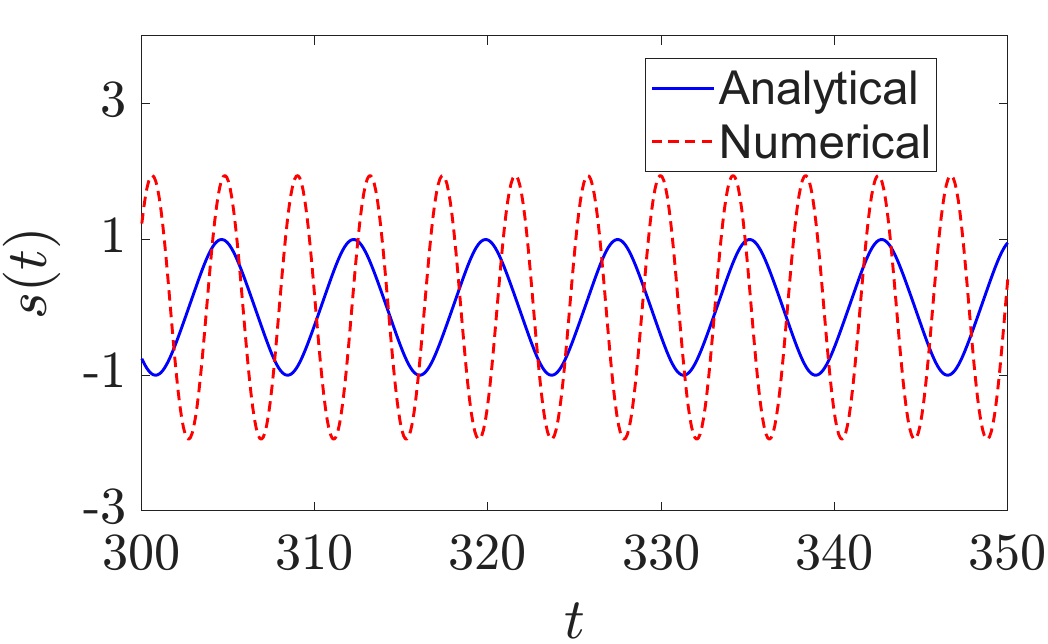}}

     \subcaptionbox{Phase space}{\includegraphics[scale=0.45]{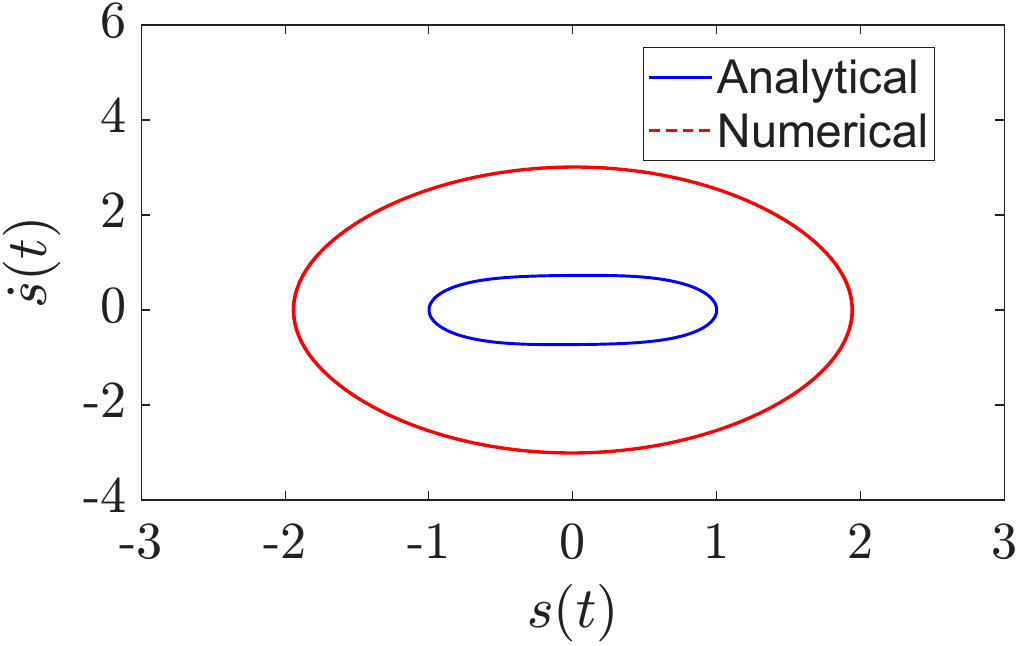}}
    \caption{Comparison of (a) time series and (b) phase portrait for sustained oscillation at $h=1.5>h_{th}$. Numerical results are obtained from Eq.\eqref{eq:effective_dynamics} along with the analytical result in Eq.\eqref{eq:final_slow_sol}. Different parameters are: $\alpha=0.3,\gamma=0.1,p=1,q=1,\omega_0=1.0,\omega_p=3.0$ and $g=150,\Omega=10$.}

    \label{fig:ana_vs_numer}
\end{figure}

\begin{figure}[h]
     \subcaptionbox{}{\includegraphics[scale=0.5]{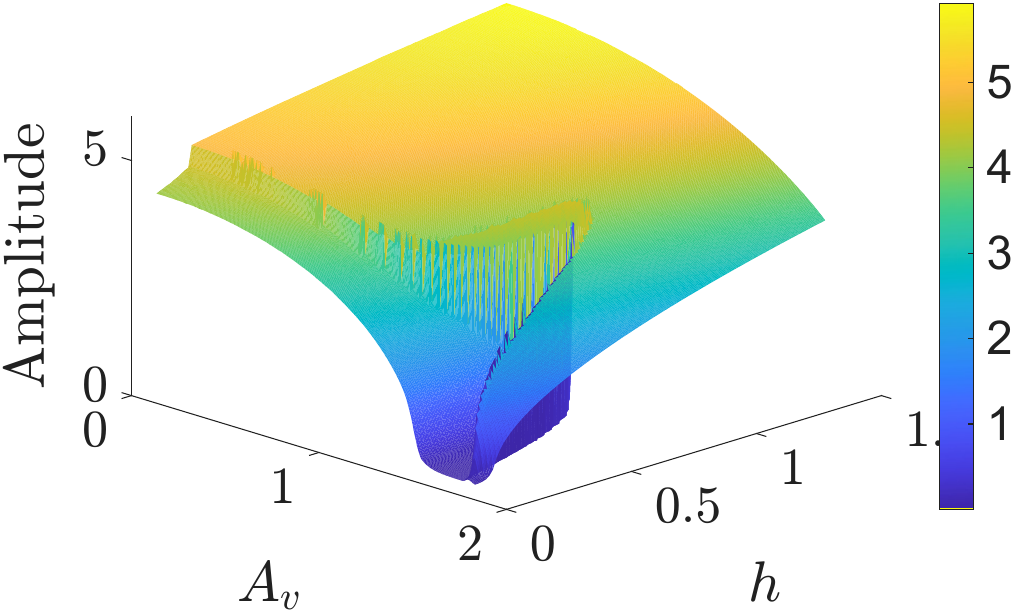}}
    \subcaptionbox{}{\includegraphics[scale=0.5]{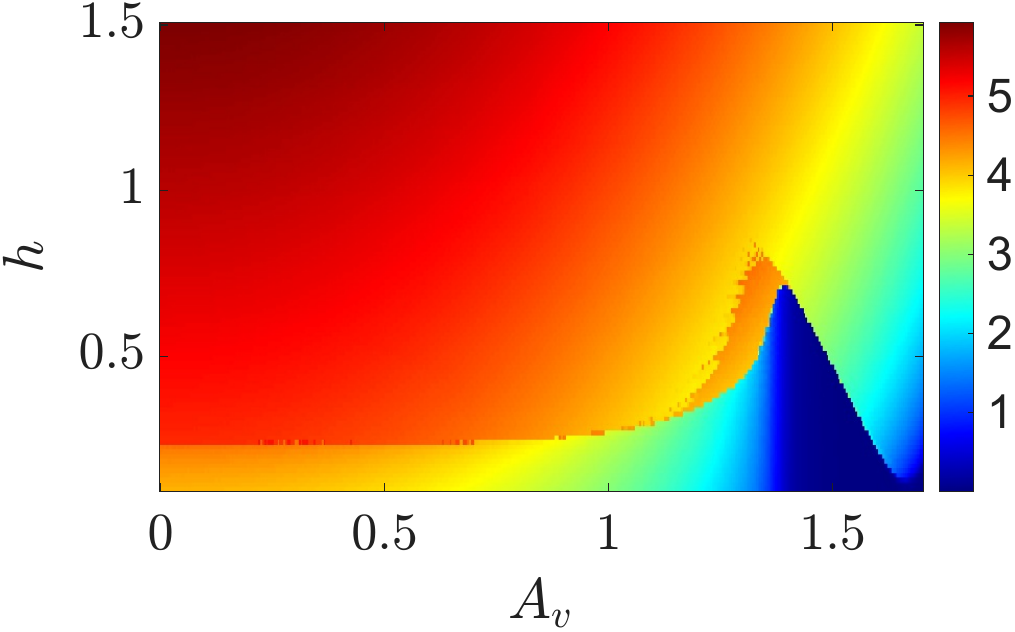}}
    \caption{Variation of amplitude in ($h-A_v$) parametric plane (a) 3D visualization (B) 2D representation by color map obtained from the effective dynamics Eq.\eqref{eq:effective_dynamics}.Different parameters are: $\alpha=0.3,\gamma=0.1,p=1,q=1,\omega_0=1.0$, and $\omega_p=3.0$.}

    \label{fig:heat_map}
\end{figure}

In the following we shall study the possibility of Hopf bifurcation as we vary the signal strength $A_v$. Through a linear stability analysis of Eq.\eqref{eq:rg_flow_amp}, we see that the origin is unstable while the second root of $a$, denoting the radius of the limit cycle, is stable. Clearly, this non-trivial root of $a$ reveals the dependence of the amplitude of the limit cycle on the fast signal strength $A_v$ through $K(A_v)$ defined in Eq.\eqref{eq:effective_damp}. The goal of using the parameter $A_v$ is to quantitatively understand how much power, proportional to \(A_v^2 \), is needed in the fast excitation to control the stability of oscillations. A relative evaluation of the effectiveness of control strategies is made possible by this representation, especially when deciding whether frequency detuning through high-frequency forcing provides an effective path to either stabilize or destabilize the system. The condition \(\left(\partial a/\partial \mu\right)_{a=r_0}=0\), gives a trivial solution \(r_0=0\) and a nontrivial amplitude

\begin{equation}
  r_0=\frac{2}{\sqrt{p}}\sqrt{K}=\frac{2}{\sqrt{p}}\sqrt{q-\frac{A_v^2}{2}} \label{eq:limit_radius} 
\end{equation}

\noindent From Eq. \eqref{eq:rg_flow_amp}, it is also clear that in the parametric space, $K=0$ represents a point where a supercritical Hopf bifurcation occurs. When $K < 0$, the origin acts as the only stable fixed point. Looking back at Eq. \eqref{eq:effective_damp}, we see that by changing the control parameter $A_v$, we can change $K$. Therefore, when $K$ becomes positive, we see that the origin is destabilized, and a stable limit cycle is created. This transition occurs at the point $K=0$ which, according to Eq. \eqref{eq:effective_damp}, happens at the threshold value of $A_v$ as well as $g$ are given by
\begin{equation}
    (A_v)_{hopf}=\sqrt{2q}~~\text{or}~~g_{hopf}=\sqrt{2q}~\Omega^2
    \label{eq:hopf_condition}
\end{equation}

The existence of a stable limit cycle is guaranteed by the condition \( A_v < \sqrt{2q} \), as can be seen from the equation above. Besides Eq.\eqref{eq:effective_freq} gives the resonance condition at sub-harmonic \(\tilde{\omega}=\frac{\omega_p}{2}\), that is, when the system oscillates with maximum amplitude, yields,

\begin{equation}
    (A_v)_{res}=\sqrt{\frac{2}{3\alpha}\left(\frac{\omega_p^2}{4}-\omega_0^2\right)}~~;~~~~\text{for}~~\omega_p>2\omega_0
    \label{eq:res_condition}
\end{equation}

The above relation has been widely used to find the peak of forcing strength $g$; we are not going to discuss it in detail, as this aspect has already been explored in a plethora of works concerning vibrational resonance, as cited in the introduction. Now, combined with Eq.\eqref{eq:threshold_para_strength} and Eq.\eqref{eq:hopf_condition}, we are now in a position to discuss our numerical results in the following section.

\section{Numerical results and discussion}
 The numerical results presented here to support our preceding analytical outcomes can be categorized into two segments: 

 $1.$ The modulation/or tuning of the limit cycle when parametric strength $h$ remains below the $h_{th}$.
 
 $2.$ In the scenario when $h>h_{th}$, i.e the onset of entrained subharmonic oscillation.We have chosen a fixed parameter set for the systematic evaluation of the results, which are given as $\alpha=0.3,\gamma=0.1,p=1,q=1,\omega_0=1,\omega_p=3.0$ and $\Omega=10$. Here we have taken $A_v=g/\Omega^2$, the strength of the signal as a control parameter, rather than treating $g$ and $\Omega$ as separate parameters. The motivation behind this lies in its experimental procedures, where the question might arises about the efficiency of tuning the system dynamics by applying additional fast forcing. Therefore, it should not be confused with the magnitude of the force $g$; rather, as the input power is proportional to $A_v^2$, which scales as $~1/\Omega^4$, indeed remains very small, since the framework permits us to choose $\Omega$ to be sufficiently large. For $\Omega=10$, $(A_v)_{hopf}=1.414$ from Eq.\eqref{eq:hopf_condition}, we keep $g=135$ for which the parametric threshold from Eq.\eqref{eq:threshold_para_strength} reads $h_{th}=0.86$, and we fix $h=0.1<h_{th}$. Now, for the chosen value of $g=135$, the amplitude of the fast signal $A_v=1.35$ which is below $(A_v)_{hopf}$, we can expect a limit cycle oscillation, which can be validated from the original Eq.\eqref{eq:original} and the effective dynamics Eq.\eqref{eq:effective_dynamics} also, see Fig. \ref{fig1}. We have followed the numerical procedure as described in \cite{landa2000vibrational} to simulate the original dynamics Eq.\eqref{eq:original}. The main idea is to extract the slow dynamics, also known as the envelope dynamics, by averaging out the fast scale. This is realized by taking the Fourier transform of the state component. In Fig.\ref{fig1}, while a sustained limit cycle oscillation can be observed, in Fig.\ref{fig2} that limit cycle is collapsed and the phase flow tends towards a fixed point when $A_v=1.5>(A_v)_{hopf}$ for $g=150$, indicating a supercritical Hopf bifurcation. Also in Figs.\ref{fig1}(c) and \ref{fig2}(c), the FFT of the original dynamics shows the presence of two major frequencies $\omega_p/2$ and $\Omega$ as expected, but in the FFT of the effective (envelope) dynamics Figs.\ref{fig1}(e) and \ref{fig2}(e),  only the peak near \( \omega_p/2 \) remains, since the fast component is effectively averaged out in the envelope formulation.\\

 Now in the second scenario, when $h>h_{th}$, is displayed in Fig.\ref{fig:h_gt_h_th}. We observe an entrained oscillation at $\omega_p/2$, despite the fact that the signal strength $A_v=1.5,$ which exceeds the Hopf threshold value $(A_v)_{hopf}=1.41$. This entertainment is confirmed by the FFT of the effective dynamics (Fig.\ref{fig:h_gt_h_th}(f)), where a sharp peak is detected at $\omega_p/2$, reflecting the dominance of the subharmonic. On the other hand, if we compare the spectral power of the original dynamics from Figs. \ref{fig2}(f) and \ref{fig:h_gt_h_th}(f), it is visible that when $h<h_{th}$, the power at $\omega_p/2$ is significantly low than when $h>h_{th}$, even though in both cases $A_v>(A_v)_{hopf}$. In the entrained regime, the limit cycle doesn't converge to a fixed point and maintains a persistent subharmonic oscillation. Also the extra peaks appear, corresponding to modulation frequencies such as \( |\Omega - \omega_0| \), \( |\Omega - \omega_p| \), and \( |\omega_p - \omega_0| \). These peaks develop due to the coexistence of multiple time scales in the direct simulation of FFT from the original dynamics. In contrast, they are absent in the effective case because there, the fast scale is systematically averaged out. This averaging process, along with the confinement perturbative expansion to the first order while neglecting the higher orders, introduces a disparity between analytical and numerical results. In Fig. \ref{fig:ana_vs_numer}, the comparison of the time series and the phase portrait of the effective dynamics Eq\eqref{eq:effective_dynamics} and the analytical result Eq.\eqref{eq:final_slow_sol} obtained from RG reveals the discrepancies that should not be overlooked. Finally, in Fig.\ref{fig:heat_map} we present a color map in the parametric plane of $A_v$ and $h$, depicting the amplitude variation of the system Eq.\eqref{eq:effective_dynamics} for the chosen values of parameters. It shows that around $A_v=1.4$ the oscillation amplitude drops down when $h$ is around 0.8, which is actually the threshold of $h$ for that particular signal strength. Above this $h$ value, the system remains in a state of high amplitude oscillation even if the value of $A_v>(A_v)_{hopf}$, maintaining the sustained oscillation at subharmonic.

\section{Conclusion}
In summary, in this article, we have demonstrated the emergence of subharmonically entrained oscillation and supercritical Hopf bifurcation in a VMD oscillator subjected to a high-frequency external driving. We have employed both the direct partition of motion (Blekhman perturbation) and the renormalization group approach to derive the effective and the slow flow dynamics of the system, which indeed lead to the condition for the onset of subharmonic threshold for entrainment and the critical value of signal amplitude at which stability change occurs through supercritical Hopf bifurcation.\\
A central focus has been given to the control parameter $A_v$, formed by the forcing strength $g$ and the high frequency $\Omega$, which governs both the effective natural frequency $\tilde{\omega}$ and the damping term $K$. According to our findings, \( A_v \) can function as an effective and independent control parameter, affecting the limit cycle's stability and the threshold for subharmonic sustained oscillation, which has been derived analytically and supported by the numerical results. Besides, the FFT analysis of the original and the effective dynamics reveals the emergence of the dominant subharmonic $\omega_p/2$ at the time of entrainment.\\
Despite the two crucial agreements of critical threshold onset and limit cycle stability, there are some numerical disparities regarding the results in Fig.\ref{fig:ana_vs_numer}, obtained from the effective dynamics and from RG analysis. These limitations imply that adding higher-order corrections in perturbation expansion and long-period averaging in numerical methods can be done to improve the approximation. Overall, this work demonstrates how useful the RG method is for capturing the interplay between fast and slow dynamics and offers a systematic understanding of fast forcing control strategies in nonlinear parametric oscillators. Our findings can be applicable to experimental systems in which slow oscillatory behavior is controlled by fast driving, in biological networks, nano-scale oscillatory systems, and natural phenomena.

\section{Acknowledgment}
 SR is deeply grateful to the Honorable Director, Prof. Dr.Satyajit Chakrabarti, for providing the facility to pursue this research work at the Institute of Engineering \& Management (IEM Kolkata), and to Prof. Dr. K.P. Ghatak for his invaluable support, encouragement, and guidance.

\bibliographystyle{unsrt}
\bibliography{references}

\end{document}